\documentclass[sigconf, natbib=false]{acmart}

\usepackage[english]{babel}
\usepackage{csquotes}
\usepackage{microtype}

\usepackage[style=numeric]{biblatex}
\graphicspath{ {./figures/} }

\renewcommand\footnotetextcopyrightpermission[1]{} 
\setcopyright{none}

\acmDOI{}

\acmISBN{}

\acmConference[Submitted for review to SIGCOMM]{}
\acmYear{2021}

\acmPrice{}

\begin{document}
\title{A Model of WiFi Performance With Bounded Latency}
\author{Bj\o rn Ivar Teigen}
\affiliation{%
  \institution{University of Oslo}
}
\email{bjornite@ifi.uio.no}
\author{Neil Davies}
\affiliation{%
  \institution{Predictable Network Solutions Limited}
}
\email{neil.davies@pnsol.com}
\author{Kai Olav Ellefsen}
\affiliation{%
  \institution{University of Oslo}
}
\email{kaiolae@ifi.uio.no}
\author{Tor Skeie}
\affiliation{%
  \institution{University of Oslo}
}
\email{tskeie@ifi.uio.no}

\author{Jim Torresen}
\affiliation{%
  \institution{University of Oslo}
}
\email{jimtoer@ifi.uio.no}

\renewcommand{\shortauthors}{Teigen et al.}

\begin{abstract}
In September 2020, the Broadband Forum published a new industry standard for measuring network quality. The standard centers on the notion of quality attenuation. Quality attenuation is a measure of the distribution of latency and packet loss between two points connected by a network path. A vital feature of the quality attenuation idea is that we can express detailed application requirements and network performance measurements in the same mathematical framework. Performance requirements and measurements are both modeled as latency distributions. To the best of our knowledge, existing models of the 802.11 WiFi protocol do not permit the calculation of complete latency distributions without assuming steady-state operation. We present a novel model of the WiFi protocol. Instead of computing throughput numbers from a steady-state analysis of a Markov chain, we explicitly model latency and packet loss. Explicitly modeling latency and loss allows for both transient and steady-state analysis of latency distributions, and we can derive throughput numbers from the latency results. Our model is, therefore, more general than the standard Markov chain methods.  We reproduce several known results with this method. Using transient analysis, we derive bounds on WiFi throughput under the requirement that latency and packet loss must be bounded.

\end{abstract}
\maketitle

\section{Introduction}
\label{sec:introduction}
In September 2020 the Broadband Forum published a new industry standard for measuring network quality \cite{Forum2020TR-452.1Requirements}. The standard is called ``Quality Attenuation Measurement Architecture and Requirements''. \textit{Quality attenuation} is a measure of the latency and packet loss performance of packet-switched networks. In this light, we revisit established modeling methodologies for the WiFi protocol because most previous work on modeling the 802.11 protocol has focused on analysis of throughput values only \cite{Bianchi2000PerformanceFunction, Engelstad2006Non-saturationPrediction, Tinnirello2010RethinkingMethodology, Tinnirello2005RevisitNetworks}. Throughput analysis can be used to calculate the WiFi link's \textit{average} latency \cite{Bianchi2005RemarksAnalysis}, but average latency is not sufficient to model Quality of Experience (QoE) \cite{1892}. 

``‘Performance’ is typically considered as a positive attribute of a service. However, a perfect service would be one without error, failure or delay, whereas real services always fall short of this ideal; we can say that their quality is attenuated relative to the ideal'' \cite{Thompson2020TowardsSystems}. The quality attenuation (abbreviated $\Delta Q$) concept has been developed through several decades of academic work \cite{bradley-1999, Davies1999End-to-endNetworks, CERN-THESIS-2013-004, 1892, Thompson2020TowardsSystems}. Thompson and Davies \cite{Thompson2020TowardsSystems} present a framework for performance management based on the notion of quality attenuation. The $\Delta Q$ framework centers on the assertion that network performance should be defined as the amount of latency and packet loss the network introduces. $\Delta Q$ can be modelled as the distribution of latency introduced by each hop along the network path, with packet loss modelled as infinite latency. \cite{Thompson2020TowardsSystems} shows how the $\Delta Q$ of a network link can be used to model application performance over that link. In particular, the tail of the latency distribution at each hop is important for the end-to-end performance of an application, especially when many hops are involved in transmitting data. Understanding the tail of the latency distribution is therefore key to understanding network performance as seen from the end-user perspective.

The model described by Bianchi \cite{Bianchi2000PerformanceFunction} is perhaps the most influential WiFi model in the literature. The analysis in \cite{Bianchi2000PerformanceFunction} is performed using steady-state analysis of a Markov chain description of the WiFi protocol. Such steady-state analysis is suitable for analyzing average throughput over long time-scales. Throughput is proportional to average latency when the system is saturated, as shown in \cite{Bianchi2005RemarksAnalysis}, and throughput and average latency therefore represent the same information about the system performance at saturation. However, for the purposes of relating WiFi performance to application-level outcomes, we need a more complete description of the latency distribution. This work presents a novel WiFi model that describes complete latency distributions and packet loss probabilities. We can describe long-term average throughput by computing the average latency, and our method is thus more general than the steady-state Markov chain analysis approach. 

This work analyzes the latency and packet loss performance of the WiFi protocol. We propose a method that explicitly models latency and packet loss. Evaluating our model requires more computational resources than a comparable Markov chain method, but we gain the ability to do transient analysis because we do not rely on the system being in a steady state. We compute throughput values from the latency results and show that our throughput values match those derived by Markov chain methods.  We then derive bounds on throughput under the requirement that latency and packet loss must be bounded. We also reproduce some know results such as the WiFi performance anomaly \cite{Heusse2003Performance802.11b}. Our model is suitable for analysis of application layer performance using the methods described in \cite{1892, Thompson2020TowardsSystems} because it accurately models the tail of the latency distribution.

Section \ref{sec:background} lays out the most relevant related work. We explain our method and its application to WiFi in section \ref{sec:method}. In section \ref{sec:results}, we validate the convergence and accuracy of our model. In section \ref{sec:latencyanalysis}, we use our model to find an upper bound on WiFi throughput with latency guarantees. We expand on the work on upper bounds in section \ref{sec:modernstandards} by exploring the impacts of several improvements to the WiFi protocol. Finally, we conclude the work in section \ref{sec:conclusion}. This work does not raise any ethical issues.

\section{Background}
\label{sec:background}

Reeve \cite{1892} shows that mean value analysis of network latency is not sufficient to model application performance. One reason why average latency does not capture the notion of performance is that network users care about how reliably an outcome is delivered on time. We require a model that can capture the risk of not delivering the desired outcome in a specified time. Mean latency is not at all sufficient to capture this risk. Consider a network that loads a website in 1 millisecond 99 out of 100 times, but once every 100th time, loading the website takes 10 seconds. The average delay of this outcome is only about 100ms. An observer monitoring this network using average values only might conclude that a 100ms load time is very reasonable, but the unpredictable behavior is likely to annoy users.

Thompson and Davies \cite{Thompson2020TowardsSystems} show how the notion of quality attenuation is related to the probability of delivering application outcomes in time.

Bianchi \cite{Bianchi2000PerformanceFunction} models the WiFi distributed coordination function (DCF) using a Markov chain, and in doing so, makes a few key simplifications. The most important simplification is to abstract away the details of delays in the model. A time-step in the model is defined by the value of the back-off counters. That is to say, the model does not separate the case in which the medium is idle from the case in which the station (STA) has to wait for another transmission to complete before the back-off counter is decremented. In other words, the time-steps are defined in terms of the model state, not how much actual time has passed. Defining the time-steps in terms of back-off counter values is a useful simplification for a Markov chain analysis, but it comes at the cost of discarding timing information. Bianchi also points this out \cite[Section IV, A]{Bianchi2000PerformanceFunction}.

Tinnirello et al. \cite{Tinnirello2010RethinkingMethodology} extend the methodology of Bianchi \cite{Bianchi2000PerformanceFunction}. Here, a Markov chain is solved for the steady-state \textit{distribution of back-off timer values}. While this approach was chosen to better model the different channel access probabilities of the Wireless Multi-Media (WMM) extension of WiFi, this method is also closer to modeling latency distributions. Tinnirello et al. deals with the distribution of back-off timers instead of simply the probability of packet transmission. Their model still makes simplifications that hide latency information because the model does not deal with differences in transmission times due to different data rates. Heusse \cite{Heusse2003Performance802.11b} shows that differences in data rates are very important for WiFi performance. A time-step in Tinnirello's model is defined as the period from the end of one transmission or collision to the end of the next transmission or collision. Thus, this model does not take into account that the size of the transmission time interval may change. That is not to say this model is incorrect, only that it lacks fidelity in modeling latency.

In \cite{Engelstad2006Non-saturationPrediction}, Engelstad et al. expand the model of \cite{Bianchi2000PerformanceFunction} to include the Access Categories of 802.11e and to handle both saturated and unsaturated networks. The model is validated by comparison to a simulation, but only throughput numbers are reported.

Youm and Kim \cite{Youm2013LatencyLANs} derive latency distributions from the model of Bianchi \cite{Bianchi2000PerformanceFunction}. Their analysis begins by assuming the system is in the steady-state and assigning the appropriate latency values to each of the transitions. This approach is similar to the one we present in that it computes  latency distributions by explicitly modelling the latency of each possible transition from one state to the next. Our work differs from that of Youm and Kim by avoiding the assumption that the system starts out in the steady state. Avoiding this requirement allows us to perform transient analysis starting from any system state.

From our review of previous work we see a lack of analysis of the latency and packet loss performance of WiFi. We address this shortcoming by developing a novel model of the 802.11 WiFi protocol that allows for the exact computation of the statistical distribution of latency and packet loss.

\section{Method}
\label{sec:method}
Our model is based on the quality attenuation framework ($\Delta Q$ framework) \cite{Thompson2020TowardsSystems}. Conceptually, quality attenuation is the delay and loss introduced by a network element. It describes how far the network element is from delivering the ``perfect service'' of zero delay and no chance of loss. Quality attenuation, denoted by $\Delta Q$, can be described by an improper random variable describing the probability distribution over the possible latency of an outcome. The variable is improper because there can be some chance that the outcome is never delivered. The possibility of an outcome never being achieved is modeled by including infinity in the domain of the random variable.

Analyzing the quality attenuation of the WiFi protocol is a matter of describing the latency of every possible path through the protocol state machine. The $\Delta Q$ framework provides the tools for adding up the contributions (the $\Delta Q$'s) from each possible path so that the complete system's latency distribution can be accurately described.

\subsection{Labeled Transition System}
\begin{figure}
    \centering
    \includegraphics[width=\linewidth]{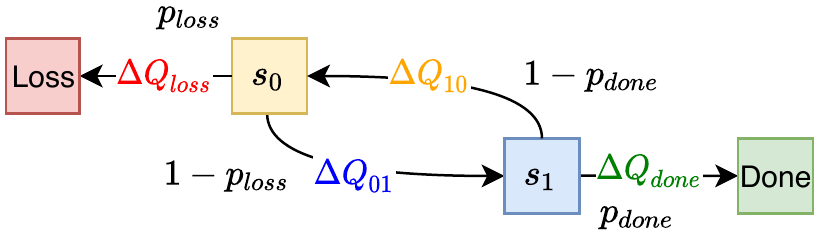}
    \caption{An example labeled transition system}
    \label{fig:lts}
\end{figure}

Our model of the WiFi protocol describes the protocol state machine as a labeled transition system. A Labeled Transition System (LTS) is a directed multigraph with labeled edges (see the example LTS in Figure \ref{fig:lts}). Nodes in the graph represent states of the protocol state machine, and edges represent transitions between the states. Each edge is labeled with a $\Delta Q$ value, which describes the distribution of time needed to complete the transition and includes the probability that the transition never terminates. Each edge also has a probability associated with it, which describes the likelihood of the system progressing through that edge conditioned on the system being the source state of the edge.

\subsection{Unrolling the labeled transition system}
In the LTS shown in Figure \ref{fig:lts}, each transition is labeled with a distribution of possible delays and a probability of that transition from the source state.

\begin{figure}
    \centering
    \includegraphics[width=\linewidth]{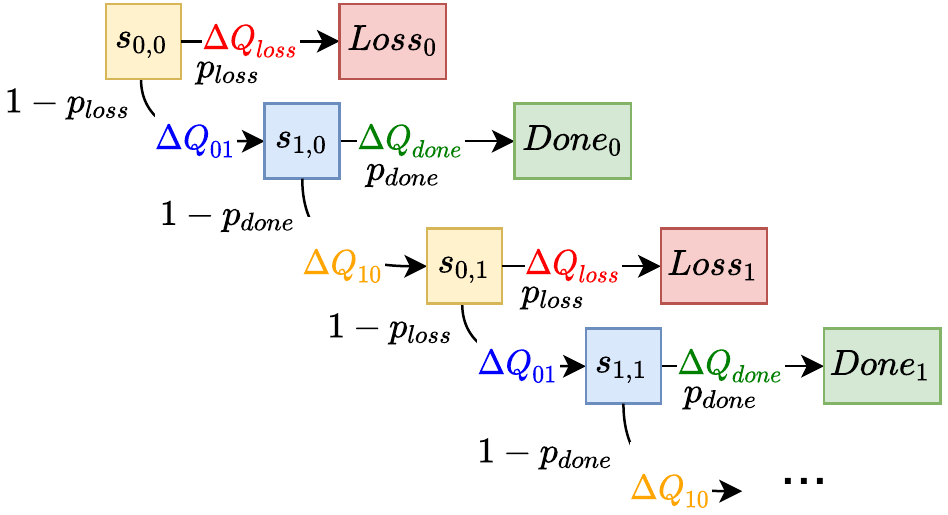}
    \caption{The unrolled labeled transition system.}
    \label{fig:tree}
\end{figure}

To make the states of the LTS Markovian, we transform the LTS so that every possible path to each state ends up in a separate copy of that state. In the unrolled LTS shown in Figure \ref{fig:tree}, new indices are added to distinguish different versions of the states from Figure \ref{fig:lts}. The cost of this transformation is a large (possibly infinitely large) increase in the size of the state space. If the LTS contains loops, this transformation will introduce infinite recursion, as shown in Figure \ref{fig:tree}. We can solve this problem by defining a maximum latency and equating any path that exceeds this limit with a loss. The WiFi protocol state machine contains no loops, and so introducing the maximum latency is unnecessary in this case, but we include the idea here to show the generality of the method.

\subsection{Composable operations}

\subsubsection{Convolution}
\begin{figure}
    \centering
    \includegraphics[width=0.8\linewidth]{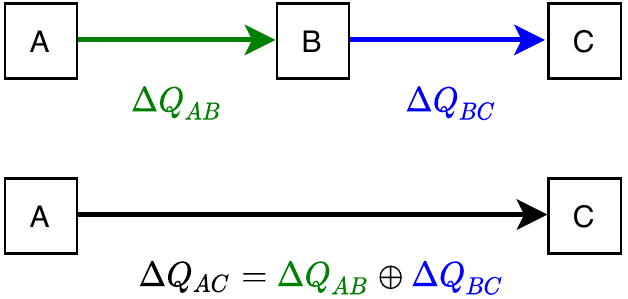}
    \caption{$\Delta Q$ convolution}
    \label{fig:convolution}
\end{figure}

Our goal is to compute the time required for the LTS to evolve from a given starting state to some state which represents the desired outcome. This calculation includes answering the question; ``Knowing the starting state, how long does it take to reach a given state?''
For states that can only be reached by a single path through the graph, we simply sum the latency contributions of each transition along that path. In the unrolled version of the LTS (see Figure \ref{fig:tree}) we have, by design, only one path to \textit{every} state. Because the latency of each edge is described by an improper random variable, the total $\Delta Q$ of a sequence of transitions can be calculated by convolving the $\Delta Q$ of each transition \cite{bradley-1999}. See Figure \ref{fig:convolution}.

\subsubsection{Mixture density}
\label{sec:mixturedensity}
\begin{figure}
    \centering
    \includegraphics[width=\linewidth]{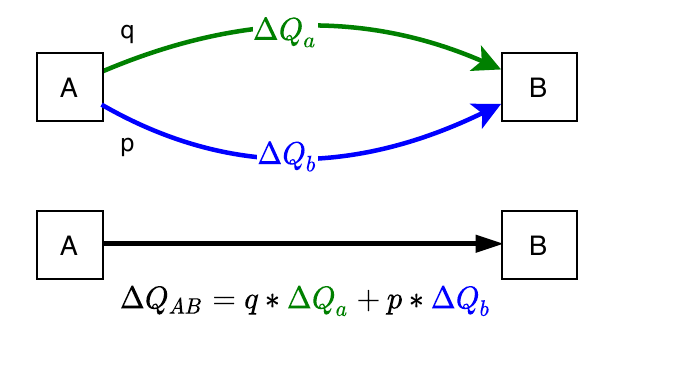}
    \caption{$\Delta Q$ mixture density}
    \label{fig:addition}
\end{figure}

When we have more than one path from the starting state to the target state, we cannot compute the arrival time distribution to the target state with convolution alone. In this case, we create a mixture distribution consisting of the latency along each possible path. The weights in the mixture distribution are determined by the probability of taking each of the paths, see Figure \ref{fig:addition}.

\begin{figure}
    \centering
    \includegraphics[width=\linewidth]{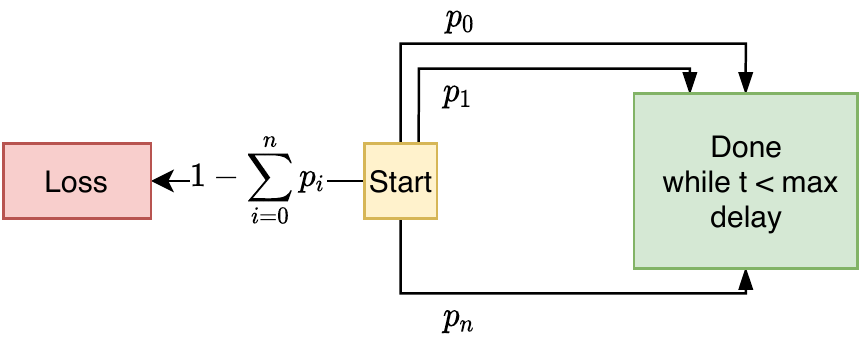}
    \caption{The reduced form of the WiFi protocol model}
    \label{fig:reducedform}
\end{figure}

There are only two possible outcomes for a packet going through the WiFi protocol stack: Successful transmission or packet loss. Consider the branches that go to the state ``Done'' in Figure \ref{fig:lts}. We unroll the LTS as illustrated in Figure \ref{fig:tree} and perform the convolution operation above on each of the paths ending in a copy of the ``Done'' state. Now, we know the latency distribution associated with each possible path to the ``Done'' state. The total latency of the done state in the not-unrolled LTS is, therefore, the mixture density formed by weighting each of the possible latency distributions that terminate in a copy of the ``Done'' state by their respective probabilities. See Figure \ref{fig:reducedform}.

\subsection{WiFi protocol background}
The WiFi protocol is ``listen before talk''. That means that a WiFi station must check that the radio frequency is idle for a certain amount of time before starting a transmission. When a WiFi station begins a transmission procedure, it first selects a random number in the range $[0, CW_{min}]$ and assign the number to a \textit{back-off counter} \cite[Section 10.2.2]{Man2013}. ``CW'' here stands for contention window. If the back-off counter is not zero, the station will wait for a single ``slot time''. If the radio frequency is sensed to be idle during the waiting period, the back-off counter is decremented by one. When the back-off counter becomes zero, the station starts transmitting. The reason for this somewhat convoluted scheme is that a station cannot listen while transmitting. This limitation means a transmitting stations cannot sense that another station is also transmitting at the same time. Therefore, collisions can not be handled by both stations interrupting their ongoing transmissions. When many WiFi stations compete for access to the frequency, collisions can waste a significant amount of time. The back-off counter mechanism was introduced to reduce the risk of collisions.

When a station is involved in a collision, the station will again choose a random value for its back-off counter. The size of the interval from which random values can be selected is a function of how many times a packet has been retried. The back-off window size doubles with each new retry of the same packet, up to a maximum value of $CW_{max}$. $max retries$ determines how many times to retry a packet before it is dropped. See Table \ref{tab:params} for the values of each parameter.

\subsection{802.11 Model details}
We represent the state of a WiFi system with $n$ competing stations as an allocation of each of the $n$ stations to a back-off counter and retry counter value. Table \ref{tab:staterep} shows an example of the state representation. Each entry in the state representation counts how many stations have a specific combination of back-off counter and retry counter values. In the example in Table \ref{tab:staterep}, two stations are at back-off counter value two after zero retries, and one station is at back-off counter value six after one retry. Note that our state representation is similar to that of Bianchi \cite[Figure 4]{Bianchi2000PerformanceFunction}.

\begin{table}
    \centering
    \begin{tabular}{c|cccccccc}
                &\multicolumn{8}{c}{Back-off counter value} \\
                & 0 & 1 & 2 & 3 & 4 & 5 & 6 &$\dots$\\ \hline
        retry 0 & 0 & 0 & 2 & 0 & 0 & 0 & 0 &$\dots$\\
        retry 1 & 0 & 0 & 0 & 0 & 0 & 0 & 1 &$\dots$\\
        $\vdots$ & & & & & & & \\
        retry $max retries$ & 0 & 0 & 0 & 0 & 0 & 0 & 0 &$\dots$\\
    \end{tabular}
    \caption{The state representation for the 802.11 model}
    \label{tab:staterep}
\end{table}

For a given state, the 802.11 protocol defines the possible transitions to a next state. Only three cases are possible:

\begin{enumerate}
    \item No stations have a back-off counter value of zero, so all stations decrease their back-off counter value by one after one slot time
    \item Exactly one station has a back-off counter value of zero. This station successfully transmits, spending the time required for transmission. The transmitting station is then finished sending its packet, and it either leaves the system or restarts the back-off procedure with a new packet. The transmitting station resets its retry counter to zero. The remaining stations hold their back-off counters constant for the duration of the transmission.
    \item More than one station has a back-off counter value of zero. This causes a collision, spending the amount of time required for the slowest of the colliding stations to transmit. All the colliding stations then increase their retry counter by one and select a random back-off counter value from the range $(0, CW_{r})$, where r is the new retry counter value. The stations not involved in the collision keep their back-off counter values constant for the duration of the collision.
\end{enumerate}

We include the minimum interval between subsequent WiFi transmissions \cite[Figure 10.4]{Man2013} in the time required for each transmission. The duration of this interval is $SIFS + \textit{Slot time}$ such that the earliest possible time for a transmission following a period with busy medium is $SIFS + 2 * \textit{Slot time}$. This simplifies the model because we do not need to keep track of whether a transmission just occurred or not. Designing the model this way makes it difficult to model the 802.11e extension of WiFi where the inter-frame space varies for traffic from different access categories. Future work will expand our model in this direction.

\section{Model evaluation, validation and comparison to existing WiFi models}
\label{sec:results}
In this section we explain how we evaluate our model, empirically test the accuracy of our model, and verify that that the model converges to the same latency distribution with each evaluation.

The state-space of our WiFi LTS model is large. Assuming the values for maximum retries, $CW_{min}$, and $CW_{max}$ from table \ref{tab:params}, the number of possible configurations of a single station is $\sum_{i=0}^{6} 2^{4+i} = 2032$. For $n$ stations, there are then $2032^{n}$ ways to assign them to back-off and retry counter values. Some of these assignments will be equivalent, but even so, evaluating the evolution of this system for all possible state configurations quickly becomes infeasible as $n$ increases. Progress has been made in solving large-scale semi-Markov models similar to the one we use here, although these models have been solved only up to the order of tens of millions of states \cite{Bradley2004Hypergraph-basedModels}. We approach the state-space explosion problem by using Monte Carlo simulation to approximate the evolution of the LTS model.

In this work we evaluate our model in two different ways; The case where all stations always has a packet to send, and the case where all stations only have a single packet to send. We label these ``Ergodic evaluation'' and ``Transient evaluation''.

\subsubsection{Ergodic evaluation}
\label{sec:ergodiceval}
When a station has either successfully transmitted it's packet, or the packet is dropped, the station immediately re-starts its back-off process. This corresponds to the saturation conditions used by Bianchi \cite{Bianchi2000PerformanceFunction}. We call this method of evaluation ``ergodic'' because it corresponds to the evaluation of an ergodic Markov chain. For this case we run the model forward from a random starting state until we have observed the outcome of $10^{4}$ packets. We arrive at the throughput numbers by first calculating the latency distribution seen by the head of line packet at each station, and then calculating throughput using equation \ref{eq:averagedelay}(see section \ref{sec:latencytothroughput}) appropriately scaled by the number of stations.

\subsubsection{Transient evaluation}
\label{sec:transienteval}
When a station has either successfully transmitted it's packet, or the packet is dropped, the station leaves the system. We record the \textit{time-to-empty}, defined as the time at which the last station leaves the system. We call this method of evaluation ``transient''  because we are essentially modelling the transient response of the system to one packet simultaneously arriving at each of $n$ stations. To compute the distribution of the measured time-to-empty, we evaluate the model starting from a random state $10^4$ times. The ability to do transient analysis of latency distributions is the main advantage of our model over steady-state analysis of Markov chains.

\subsection{Rate of convergence}
We now establish empirical results for the rate of convergence of our Monte Carlo simulations for both the ergodic and the transient evaluation method. For these experiments we use the parameters in table \ref{tab:params}, and 5 competing stations. This analysis does not confirm that the results produced are correct, but it shows how consistent the results are across different runs of the Monte Carlo simulation. We have chosen to look at the convergence of the 90\textsuperscript{th} percentile of latency because we are interested in accurately characterizing the tail of the latency distribution. Since events in the tail of the distribution are by definition rare, it takes longer for the 90\textsuperscript{th} percentile to converge. Therefore, these results are stronger than showing convergence of the mean latency.

\begin{figure}
    \centering
    \includegraphics[width=\linewidth]{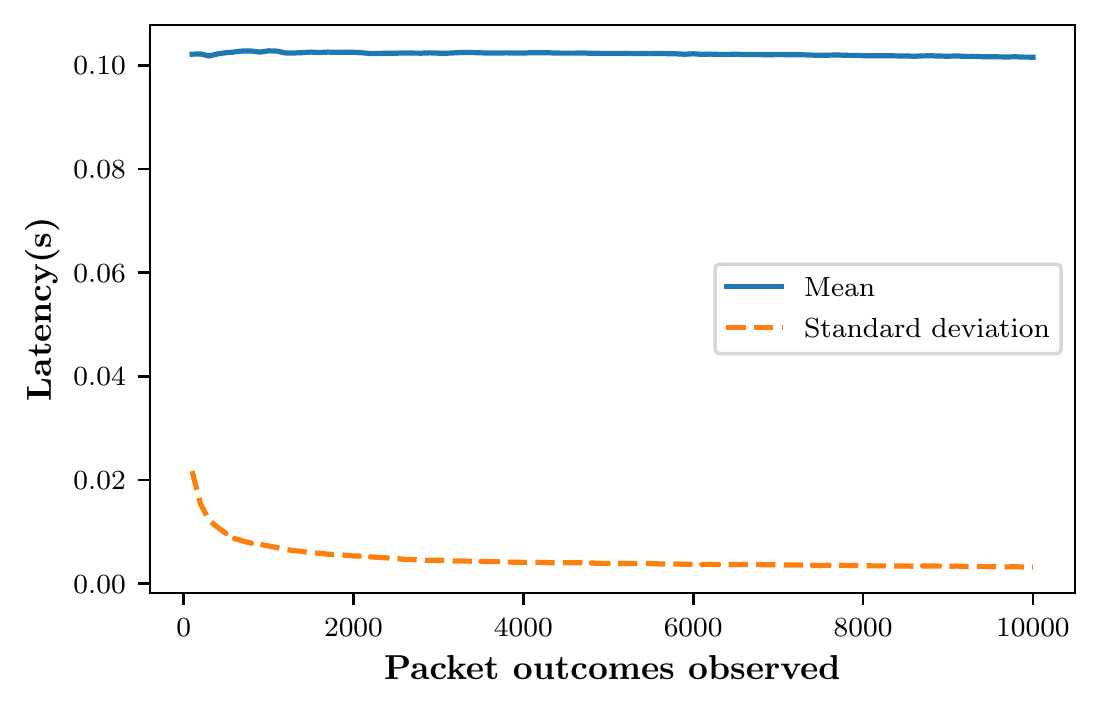}
    \caption{Convergence of the 90\textsuperscript{th} percentile latency estimate as a function of number of packet outcomes observed for the ergodic evaluation method}
    \label{fig:ergodicconvergence}
\end{figure}

Figure \ref{fig:ergodicconvergence} shows the distribution of the 90\textsuperscript{th} percentile latency as a function of the number of packet outcomes observed for the ergodic evaluation method described in section \ref{sec:ergodiceval}. We ran the simulation 1000 times to compute the distributions of the results.

\begin{figure}
    \centering
    \includegraphics[width=\linewidth]{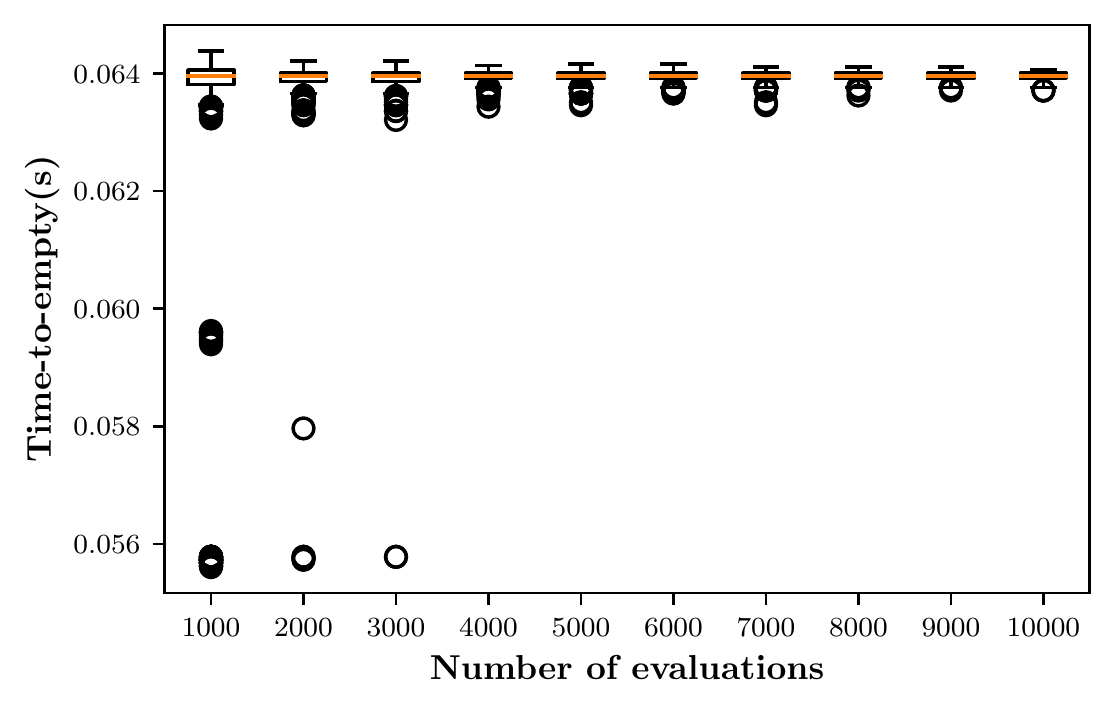}
    \caption{Convergence of the 90\textsuperscript{th} percentile time-to-empty estimate as a function of number of evaluations observed for the transient evaluation method}
    \label{fig:transientconvergence}
\end{figure}

Figure \ref{fig:transientconvergence} shows the distribution of 90\textsuperscript{th} percentile time-to-empty as a function of the number of evaluations for the transient evaluation method described in section \ref{sec:transienteval}. We ran 1000 separate simulations starting from a random state $k*1000$ times in each simulation for $k$ from 1 to 10, and recorded the 90\textsuperscript{th} percentile time-to-empty for each of the simulations.

We conclude that the 90\textsuperscript{th} percentile of the distribution converges with a high probability for both the ergodic and the transient evaluation method with the amount of packet outcomes or evaluations chosen.

\subsection{Comparison to existing models}
\begin{figure}
    \centering
    \includegraphics[width=\linewidth]{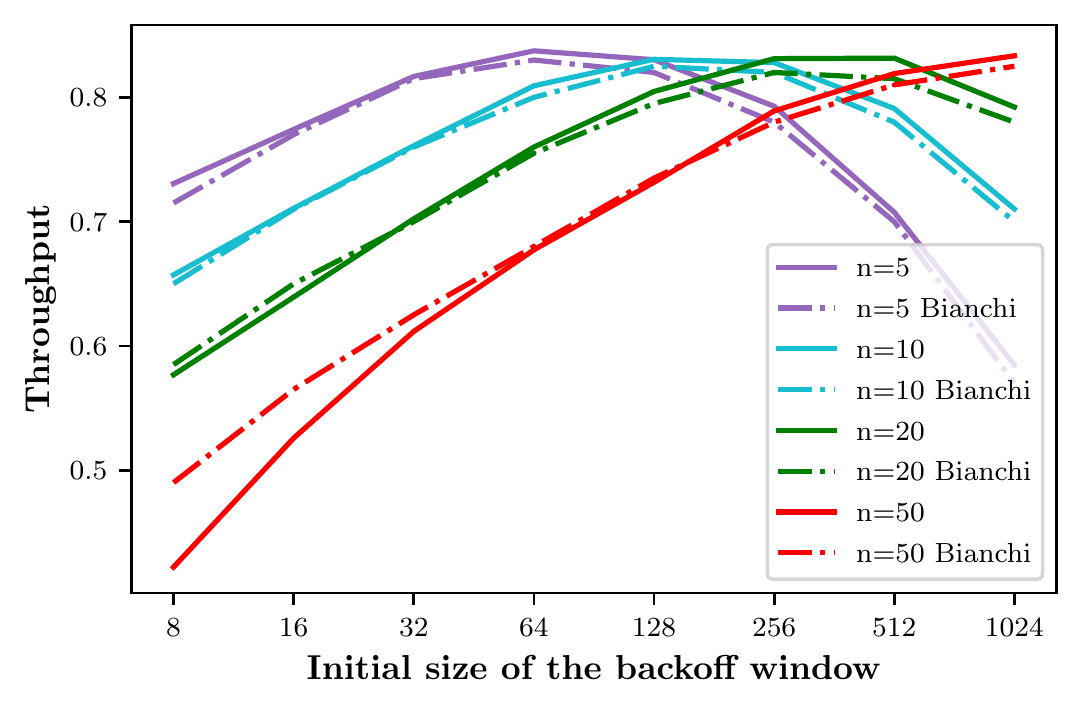}
    \caption{System throughput as a function of initial back-off window size.}
    \label{fig:ergodic}
\end{figure}

To replicate the results of Bianchi \cite{Bianchi2000PerformanceFunction} we evaluate our model using the ergodic method described in section \ref{sec:ergodiceval}. We perform the evaluation using the same parameters as Bianchi \cite[Table 2]{Bianchi2000PerformanceFunction}, shown in table \ref{tab:params}. Figure \ref{fig:ergodic} shows results for total system throughput as a function of initial back-off window size in the ergodic case, compared to results from \cite[Figure 9]{Bianchi2000PerformanceFunction}. We consider these results sufficiently close to those of \cite{Bianchi2000PerformanceFunction}, which demonstrates that our model accurately describes a saturated WiFi system for a set of different system parameters.

\begin{table}
    \centering
    \begin{tabular}{c|c}
         Parameter & Value \\
         \hline
         Slot time ($\mu s$) & 50 \\
         SIFS ($\mu s$) & 28 \\
         DIFS ($\mu s$) & 128 \\
         PHY Header (bits) & 128\\
         MAC Header (bits) & 272\\
         ACK ($\mu s$) & PHY Header + 14*8/base rate \\
         Base rate (Mbit/s) & 1 \\
         $CW_{min}$ exponent & 4 \\
         $CW_{min}$ & 15 \\
         $CW_{max}$ & 1023 \\
         $max retries$ & 6 \\
         packet size & 1023 \\
         Back-off window size & $2^{CW_{min}\text{ exponent} +\text{Number of retries}}$
    \end{tabular}
    \caption{Parameters used for comparison to the model of Bianchi\cite{Bianchi2000PerformanceFunction}}
    \label{tab:params}
\end{table}

\section{Analysis of latency in the WiFi protocol}
\label{sec:latencyanalysis}
In this section we explore the relation between latency and throughput in the WiFi protocol. First we show how latency and throughput are related in the ergodic case, and show that latency for the ergodic case grows very quickly as the number of competing stations increases. We then discuss the notion of an upper bound on throughput under the condition that latency and packet loss must be bounded.

\subsection{Relating latency to throughput}
\label{sec:latencytothroughput}
Time is related to throughput as shown in equation \ref{eq:throughputtolatency}, where $T$ is throughput in packets per second, $N$ is the number of packets sent in some interval, and $D$ is the duration of that interval (in seconds). If we also know the average packet size, for instance in bytes, we can calculate the average throughput in Mbit/s.

\begin{equation}
    \label{eq:throughputtolatency}
    T = \frac{N}{D}
\end{equation}

In our model we record the delay of each packet, and so we do not directly measure $D$ in equation \ref{eq:throughputtolatency}. Assuming the interval $D$ ends with the transmission of a packet, and that there was no idle time which did not count towards the delay of any of the recorded packets, we can calculate $D$. Consider the packets from a single station which sends packets back-to-back, as is true in the ergodic case. We denote the latency of packet $i$ by $d_{i}$, and assert $D = \sum_{i=0}^{N}d_{i}$. Observe that this means throughput is the inverse of the average per-packet delay, as shown in equation \ref{eq:averagedelay}.

\begin{equation}
    \label{eq:averagedelay}
    \frac{1}{T} = \frac{1}{N}\sum_{i=0}^{N}d_{i}
\end{equation}

Note that we only consider packets that are not lost, or else the sum of delays would be infinite. This is correct because lost packets do not contribute to the throughput. Lost packets will, however, increase the average latency for the packets that are not lost. This is also in accordance with the method used by Bianchi to calculate mean latency from throughput values \cite{Bianchi2005RemarksAnalysis}.

\subsection{Analyzing latency under saturation load}
\begin{figure}
    \centering
    \includegraphics[width=\linewidth]{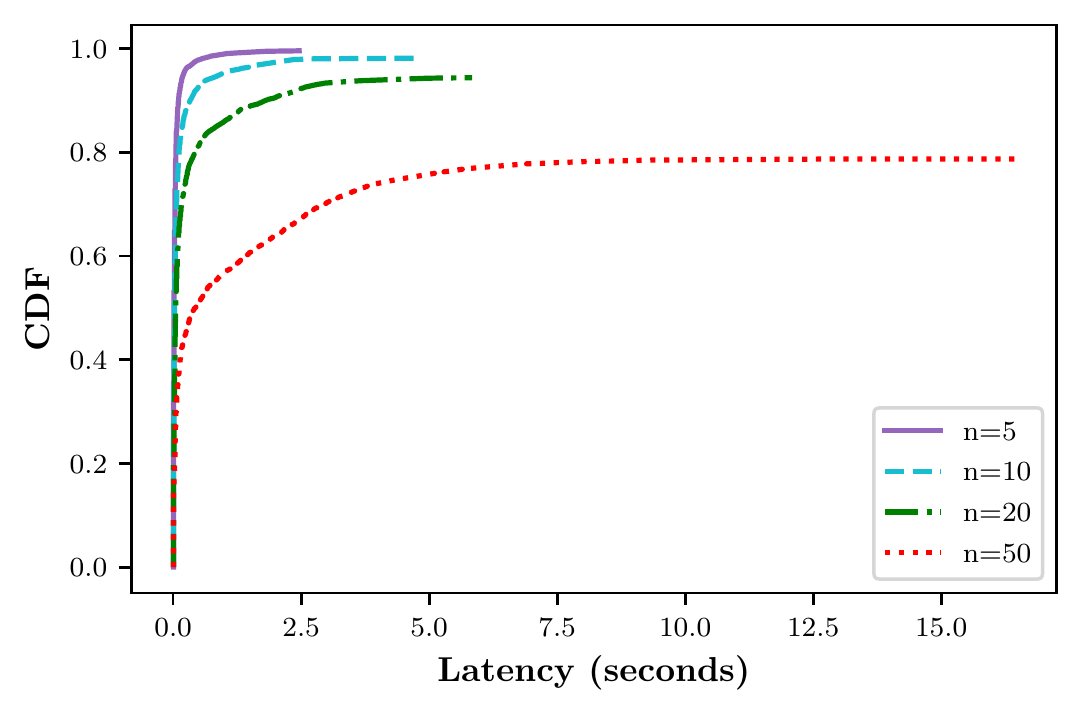}
    \caption{CDF for latency and packet loss with initial back-off window size of 8}
    \label{fig:saturationlatency}
\end{figure}

\begin{figure}
    \centering
    \includegraphics[width=\linewidth]{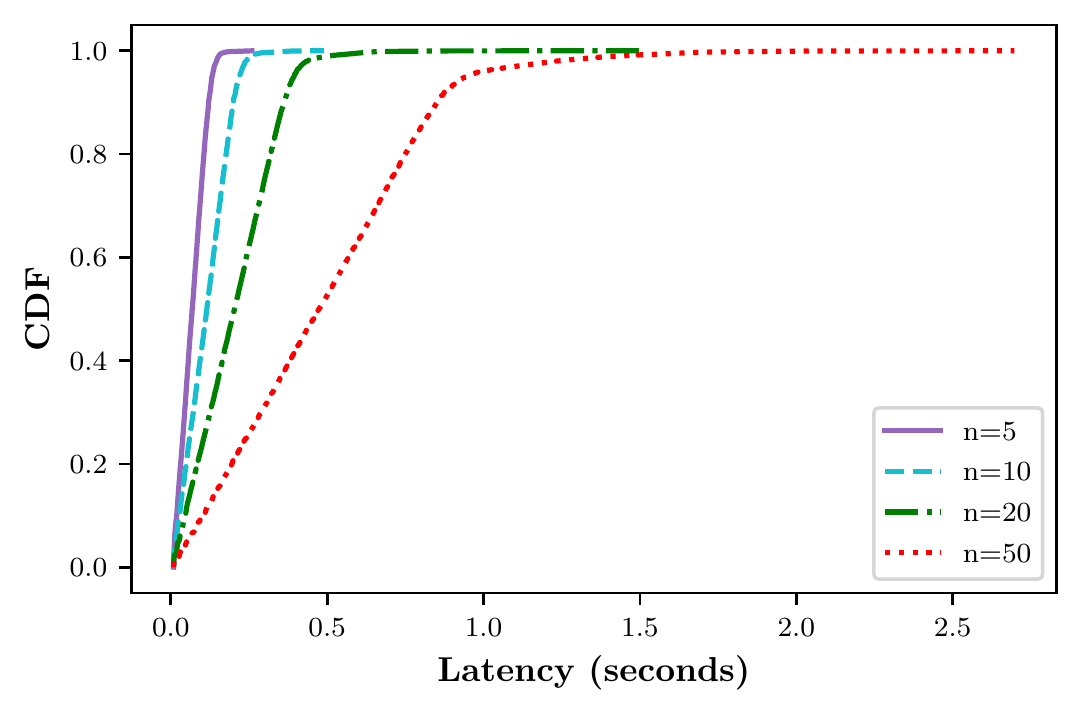}
    \caption{CDF for latency and packet loss with initial back-off window size of 1024}
    \label{fig:saturationlatencycw1024}
\end{figure}

We now proceed to investigate latency and packet loss performance in the ergodic case.
Figures \ref{fig:saturationlatency} and \ref{fig:saturationlatencycw1024} show the latency and packet loss performance of the WiFi DCF for different back-off timer values and different numbers of stations. We read packet loss values in figures \ref{fig:saturationlatency} and \ref{fig:saturationlatencycw1024} by observing how far the maximum of each CDF is from 1 on the y-axis. The quality attenuation found in these experiments is so large, especially for a high number of stations, that we argue the throughput results of Figure \ref{fig:ergodic} are of little practical use. Even though total system throughput is very close to the theoretical optimum for the 50-station case with a back-off window size of 1024, the vast majority of user applications will not perform well when running over a network with this much latency and packet loss. Interactive applications such as gaming and video conferencing obviously cannot function well with this much latency, and TCP throughput is severely affected by loss rates as high as in the 50-station case, as shown by Padhye et al. \cite{Padhye2000ModelingValidation}. Our results are consistent with those of Youm and Kim \cite{Youm2013LatencyLANs}. Note that the results show the latency of a head-of-line packet, and so queuing delays and potential packet loss due to full buffers will come in addition to the delays shown here.

Existing WiFi models mostly evaluate performance under the assumption that the system is in the steady-state and that the system is saturated \cite{Bianchi1996PerformanceLANs, Tinnirello2010RethinkingMethodology, Tinnirello2005RevisitNetworks, Youm2013LatencyLANs}. The results presented here, along with those of \cite{Youm2013LatencyLANs}, shows that the latency of WiFi under these conditions is very large. A more complete way of modeling WiFi performance is needed. We therefore argue for a different perspective on WiFi performance modeling and optimization. Instead of looking for the system parameters that will give the best throughput, or the highest system utilization, we should look for the system parameters most likely to deliver good Quality of Experience with typical end-user applications. The main drivers of QoE, as argued in \cite{1892}, are latency and packet loss. It is therefore crucial that our models accurately capture latency and packet loss performance under realistic conditions.

\subsection{Establishing an upper bound on time-to-empty}
We now compute throughput bounds under the condition that latency and packet loss must be kept bounded. We consider bounded latency and packet loss to be the absolute minimum requirement for a good user experience.

The worst-case scenario for a WiFi system with $n$ stations is that all $n$ stations begin their back-off procedure at the same time. We can think of this as all stations being maximally correlated, or as having worst-case correlation between the stations. This scenario has the highest risk of collisions and will therefore lead to the longest possible time-to-empty. We investigate this worst-case correlation scenario to establish an upper bound on the time-to-empty of a WiFi system with $n$ stations.

\begin{figure}
    \centering
    \includegraphics[width=\linewidth]{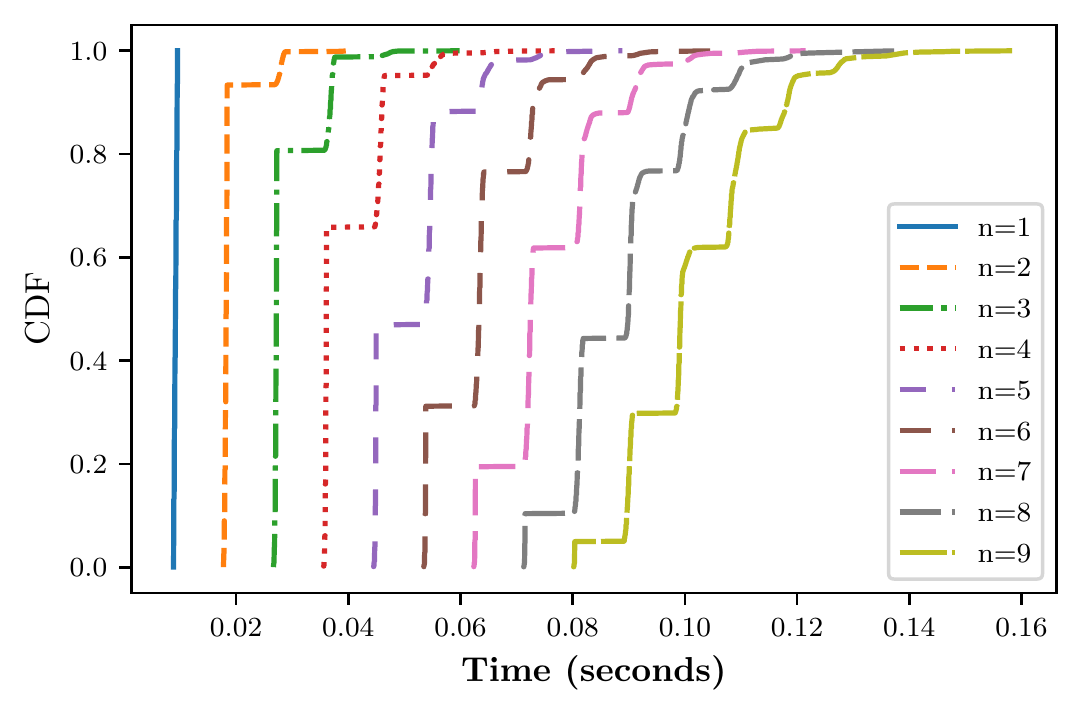}
    \caption{Time-to-empty for for $n$ competing stations starting back-off procedures at the same time}
    \label{fig:time-to-empty}
\end{figure}

The time-to-empty of an 802.11b WiFi link is shown in Figure \ref{fig:time-to-empty} for one to nine stations. To make the results comparable to those of Bianchi \cite{Bianchi2000PerformanceFunction} we use a channel rate of 1 Mbit/s. The time-to-empty represents the time from all stations simultaneously initiate the back-off process until all stations have completed the transmission of or dropped a single packet. To calculate the distribution of time-to-empty we evaluate the model using the transient evaluation method described in section \ref{sec:transienteval}

\subsection{Finding the maximum system throughput with bounded latency and packet loss}
A well-known result from queuing theory states that if, on average, packet arrivals to an unbounded queue occur more frequently than departures from the queue, then the queue length grows toward infinity. Thus, to avoid unbounded latency growth (or packet loss when queues are not infinitely large), the packet arrival rate must be smaller than the service rate. This relationship is expressed by the inequality in equation \ref{eq:arrivals}, where $\lambda$ is the mean arrival rate in packets/second, and $E[s]$ is the mean service time per packet (also in seconds).

We are now ready to find the upper bound on throughput. Our logic is as follows: If the arrival rate is slower than the worst-case service rate, then we know that latency is bounded. Using the mean service time we calculate an upper bound on the packet arrival rate by setting $\lambda = \frac{1}{E[s]}$.

\begin{equation}
\label{eq:arrivals}
    \lambda < \frac{1}{E[s]}
\end{equation}

\begin{figure}
    \centering
    \includegraphics[width=\linewidth]{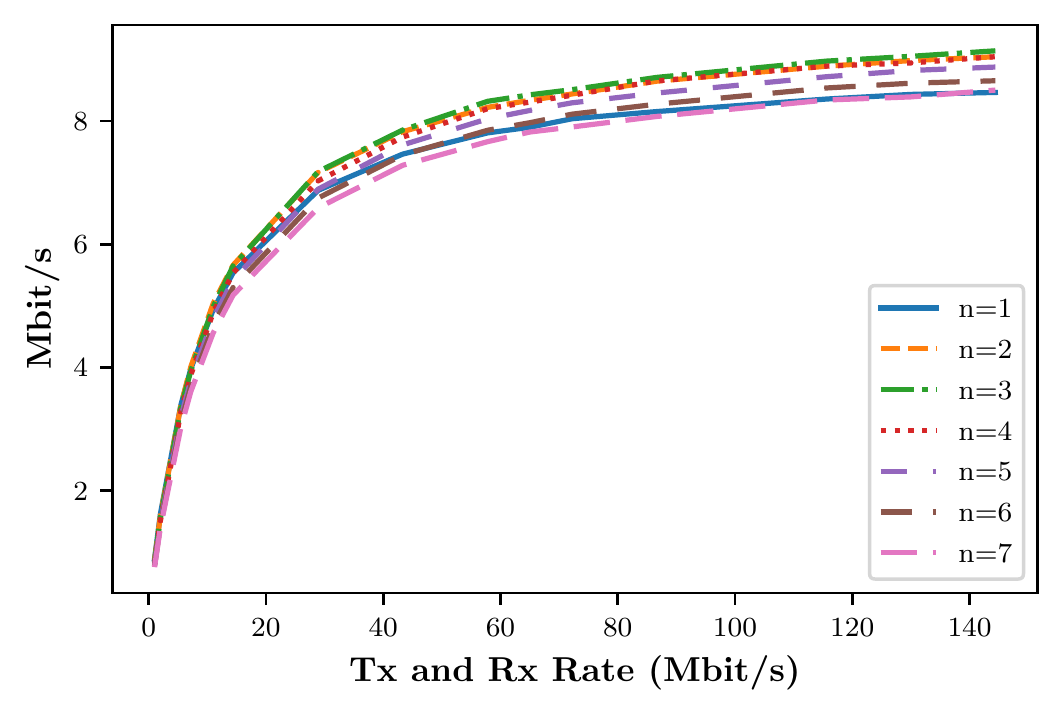}
    \caption{Upper bound on total system throughput for $n$ competing stations with bounds on latency and packet loss}
    \label{fig:roso}
\end{figure}

\begin{figure}
    \centering
    \includegraphics[width=\linewidth]{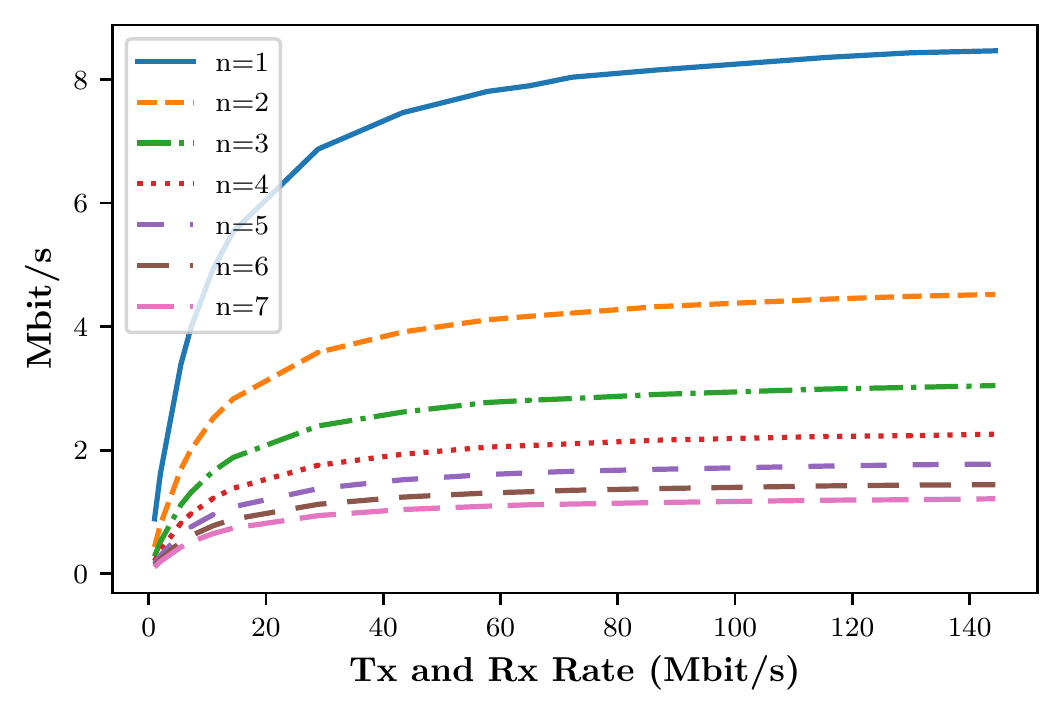}
    \caption{Upper bound on throughput per station for $n$ competing stations with bounds on latency and packet loss}
    \label{fig:rosopersta}
\end{figure}

Because the time-to-empty represents the time for the system to process one packet from each station, the upper bound on throughput is reached when one packet arrives at each of the stations every mean time-to-empty. The throughput in Mbit/s can then be calculated, assuming we know the PHY rate and the packet size. The upper bound on throughput is shown in Figure \ref{fig:roso} for one to seven stations. The throughput shown in Figure \ref{fig:roso} is total system throughput, so to compute the per-station throughput, we must divide by the number of stations. The per station throughput is shown in Figure \ref{fig:rosopersta}. These results use the parameters shown in table \ref{tab:params}, which are the same as those in Bianchi's analysis\cite{Bianchi2000PerformanceFunction}.
We now have a tool for guiding the design and configuration of WiFi networks. The upper bound on throughput gives us a way to know whether a given WiFi configuration can support a certain set of applications without building queues. We can potentially use this to inform queuing and scheduling algorithms about the available capacity of the WiFi link so that large delays due to unnecessary congestion can be avoided. In particular, our results show that WiFi performance is very sensitive to the number of simultaneously active stations on a channel. Several methods for improving WiFi performance by reducing the number of simultaneously active stations have been reported. Saeed et al. \cite{Saeed2018IfChanges} proposed a token-based WiFi scheduling algorithm for reducing contention overhead. Maity et al. \cite{Maity2017TCPSolution} proposed a WiFi scheduler for TCP downloads which reduces the number of different stations transmitting ACKs to an access point at the same time. Channel planning and transmit power management can also reduce the number of concurrently active stations and thus improve WiFi performance \cite{Akella2007Self-ManagementDeployments, DenHartog2017ABlocks}. We believe our model can help inform further work on WiFi optimization.

\section{Exploring modern WiFi standards}
\label{sec:modernstandards}
Readers familiar with WiFi performance might object that the throughput bounds presented above are too strict. Indeed, WiFi networks exist today with much greater throughput performance. In this section we explore some of the improvements that have been made to the WiFi protocol, and investigate how each improvement affects the throughput bounds.

This section explores the impact of various protocol features introduced in 802.11n and its accompanying amendments. We compute the impact on the throughput bound of Request-to-send/Clear-to-send (RTS/CTS) in section \ref{sec:rtscts} and of packet aggregation in section \ref{sec:aggregation}. We also reproduce the ``WiFi performance anomaly'' first reported by Heusse et al. \cite{Heusse2003Performance802.11b} in section \ref{sec:anomaly}.

\subsection{An 802.11n baseline}
\begin{table}
    \centering
    \begin{tabular}{c|c}
         Parameter & Value \\
         \hline
         Slot time ($\mu s$) & 9 \\
         SIFS ($\mu s$) & 10 \\
         DIFS ($\mu s$) & 28 \\
         PHY Header ($\mu s$) & 24\\
         MAC Header (bits) & 272\\
         ACK ($\mu s$) & PHY Header + 14*8/base rate \\
         Basic rate set (Mbit/s) & [1, 2, 5.5, 11, 24] \\
         $CW_{min}$ exponent & 4 \\
         $CW_{min}$ & 15 \\
         $CW_{max}$ & 1023 \\
         $max retries$ & 6 \\
         packet size & 1023 \\
         Back-off window size & $2^{CW_{min}\text{ exponent} +\text{Number of retries}}$
    \end{tabular}
    \caption{Parameters used throughout section \ref{sec:modernstandards}, unless otherwise specified. These represent the default parameters of 802.11n.}
    \label{tab:modernparams}
\end{table}

\begin{figure}
    \centering
    \includegraphics[width=\linewidth]{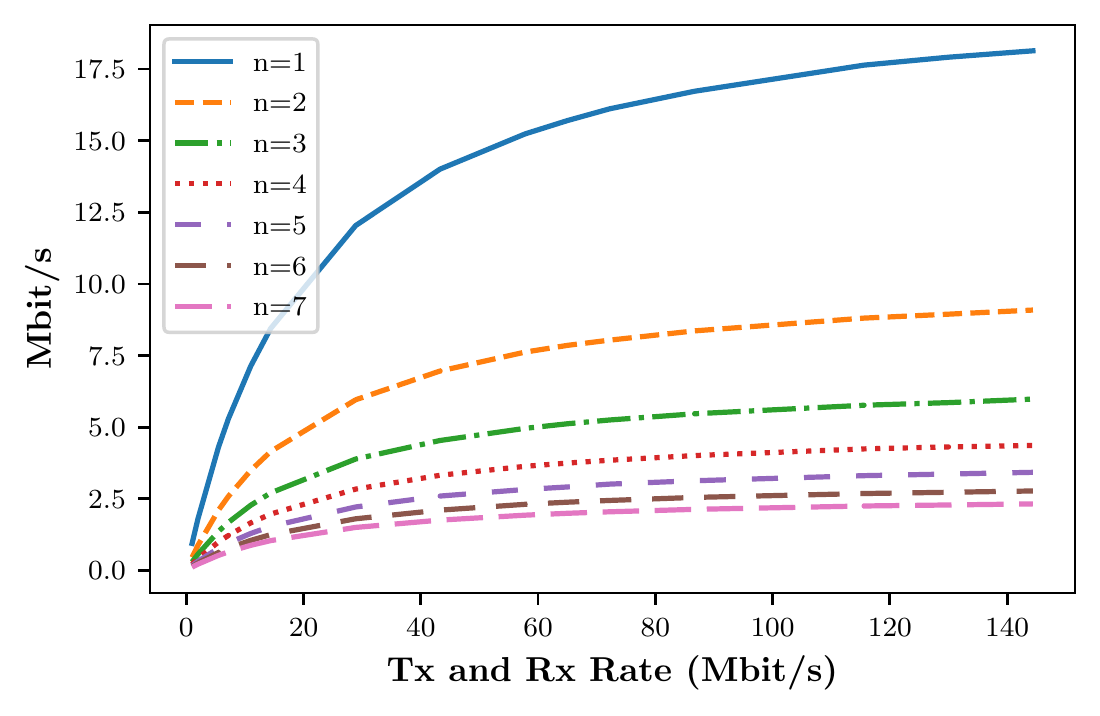}
    \caption{Upper bound on per station throughput for $n$ competing stations using 802.11n parameters (see table \ref{tab:modernparams})}
    \label{fig:rosopersta80211n}
\end{figure}

Figure \ref{fig:rosopersta80211n} shows the upper bound on per-station throughput assuming throughput is fairly divided and latency and packet loss is bounded. Comparing to Figure \ref{fig:rosopersta}, we see very similar behavior. However, as expected the throughput is significantly higher using the 802.11n parameters compared to those of \cite{Bianchi2000PerformanceFunction}. The results in Figure \ref{fig:rosopersta80211n} will serve as a baseline comparison for the protocol features explored in this section.

\subsection{RTS/CTS}
\label{sec:rtscts}
The request-to-send, clear-to-send mechanism in WiFi introduces an extra handshake between sender and receiver. Before a data packet is transmitted, the sender transmits a RTS packet. Upon hearing the RTS packet, the receiver responds with a CTS packet. If the sender hears the CTS packet, the data packet is transmitted. Because the RTS and CTS packets are small, and therefore take little time to transmit, the introduction of the extra handshake can reduce the amount of time spent transmitting whenever a collision occurs. The RTS/CTS mechanism also reduces the impact of hidden node problems.

We can model the RTS/CTS mechanism by increasing the time to complete each transmission by the time required to perform the RTS/CTS handshake. If a collision occurs, the elapsed time is decreased, because the collision is detected by all involved stations when they fail to receive a CTS frame. Because we do not consider hidden nodes in this work, RTS/CTS can only reduce the amount of time spent waiting for colliding transmission to cease.

\begin{figure}
    \centering
    \includegraphics[width=\linewidth]{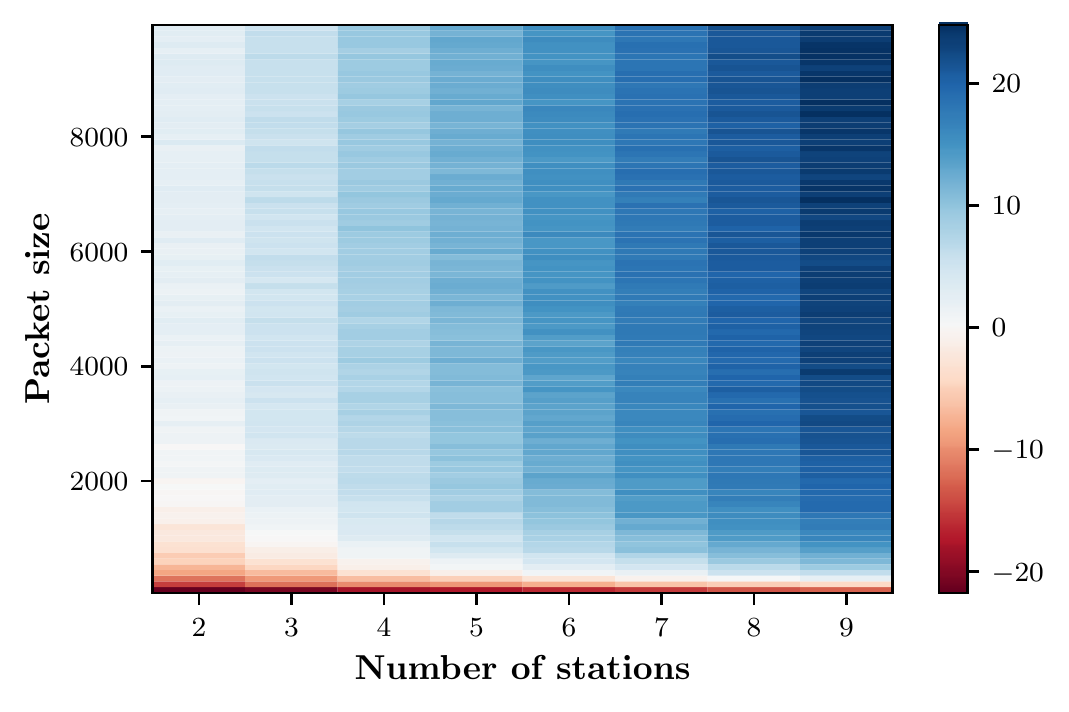}
    \caption{Heatmap showing the percent impact of enabling RTS/CTS as a function of number of stations and packet size. Positive values (towards top right corner) indicate that enabling RTS/CTS increases the upper bound on throughput. The parameters are those listed in Table \ref{tab:params}.}
    \label{fig:RTSCTSvspacketsize}
\end{figure}

We now compare our RTS/CTS results to those of \cite{Bianchi2000PerformanceFunction, Tinnirello2005RevisitNetworks}. We calculate the mean time-to-empty for a WiFi system with and without RTS/CTS enabled. We use WiFi parameters equal to those of \cite{Bianchi2000PerformanceFunction} (see Table \ref{tab:params}), and vary the number of stations and the packet size. The packet size is increased in steps of 100 bytes from 100 to 9900 bytes. Figure \ref{fig:RTSCTSvspacketsize} shows the percentage change in the upper bound on throughput from enabling RTS/CTS. Positive values indicate that the system with RTS/CTS enabled allows for greater throughput. Our results show a trend similar to those found in \cite{Bianchi2000PerformanceFunction, Tinnirello2005RevisitNetworks}, but our results indicate that RTS/CTS should be enabled for a smaller packet size and for a smaller number of stations compared to the results of \cite{Bianchi2000PerformanceFunction, Tinnirello2005RevisitNetworks}. Whereas Bianchi\cite{Bianchi2000PerformanceFunction} sets the threshold for enabling RTS/CTS in a system with five stations running at 1Mbit/s at 3160 bytes, and Tinnirello et al. \cite{Tinnirello2005RevisitNetworks} sets the threshold at 800 bytes, our results indicate that RTS/CTS should be enabled for packets above 600 bytes.

\begin{figure}
    \centering
    \includegraphics[width=\linewidth]{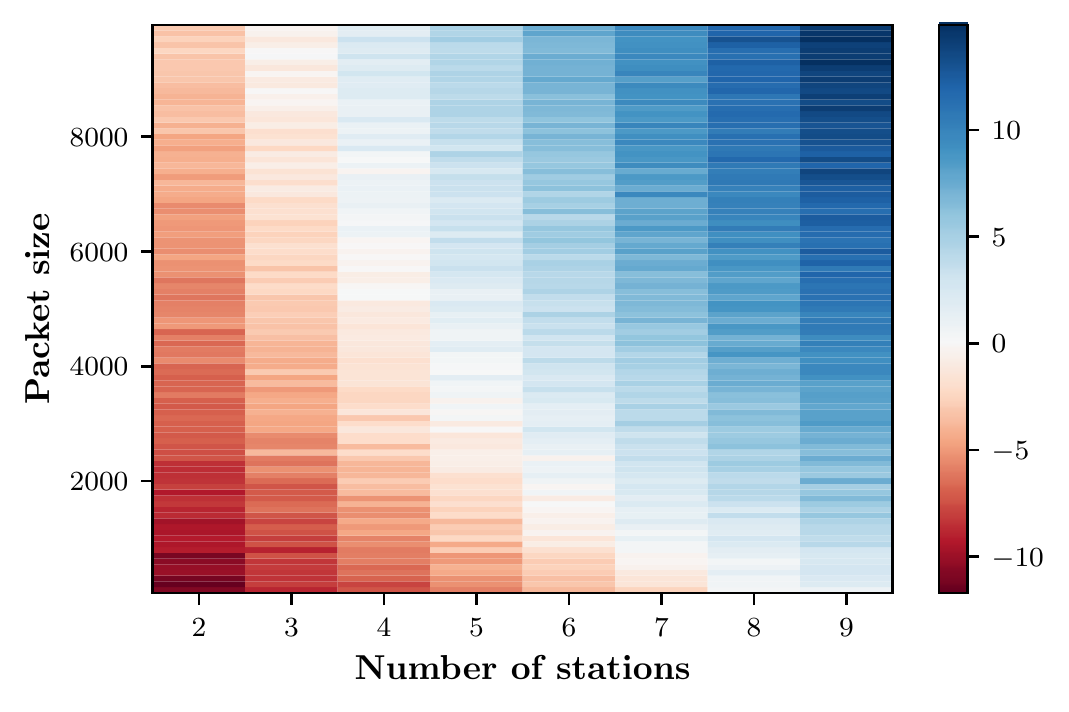}
    \caption{Heatmap showing the percent impact of enabling RTS/CTS as a function of number of stations and packet size. Positive values (towards top right corner) indicate that enabling RTS/CTS increases the upper bound on throughput. The parameters are those listed in Table \ref{tab:modernparams}. The channel rate is 144 Mbit/s.}
    \label{fig:RTSCTSvspacketsizeBaseline}
\end{figure}

Figure \ref{fig:RTSCTSvspacketsizeBaseline} shows the impact of enabling RTS/CTS for 802.11n WiFi using the parameters listed in Table \ref{tab:modernparams} and a channel rate of 144 Mbit/s. Comparing to Figure \ref{fig:RTSCTSvspacketsize}, it is clear that we are more at risk of negatively impacting the system throughput in this case, because the overhead of the RTS/CTS handshake is relatively larger when the channel rate is high. Figure \ref{fig:time-to-empty-comparison} compares the time-to-empty CDF with and without RTS/CTS enabled for 5 stations and a packet size of 1023. According to the results shown in Figure \ref{fig:RTSCTSvspacketsizeBaseline}, enabling RTS/CTS reduces total system throughput using these parameters. Figure \ref{fig:time-to-empty-comparison} clearly shows that enabling RTS/CTS reduces the likelihood of high latency, at the cost of added overhead. Increasing predictability at the cost of higher minimum latency may be a desirable trade-off for jitter-sensitive applications, even if the total system throughput decreases.

\subsection{Packet aggregation}
\label{sec:aggregation}
Packet aggregation is a mechanism by which several higher-layer packets (here typically IP packets) are transmitted together over a link, without individual MAC-layer headers. Packet aggregation increases the total throughput by reducing the MAC-layer overhead because the number of transmit opportunities required to send a given number of IP packets is reduced.
The 802.11 protocol describes the following two types of packet aggregation: Mac Service Data Unit (MSDU) aggregation and Mac Protocol Data Unit (MPDU) aggregation. To take advantage of either aggregation mechanism, the station must have buffered several packets with the same destination address.  The WiFi protocol imposes a limit on the time of each transmit opportunity. When this time expires, the station must perform the back-off mechanism.

\subsubsection{A-MSDU}
Mac Service Data Unit aggregation works by grouping several IP-layer packets into a single Mac-layer packet for the WiFi transmission. This grouping reduces the protocol overhead, both in terms of transmission time and waiting time due to the back-off mechanism. Implementing this kind of aggregation in our model is very straightforward, because increasing the packet size parameter is sufficient. In 802.11n, the maximum A-MSDU size is 7935 octets. Using this packet size, we arrive at the maximum stable throughput per station shown in Figure \ref{fig:rosoperstaAMSDU}.

\begin{figure}
    \centering
    \includegraphics[width=\linewidth]{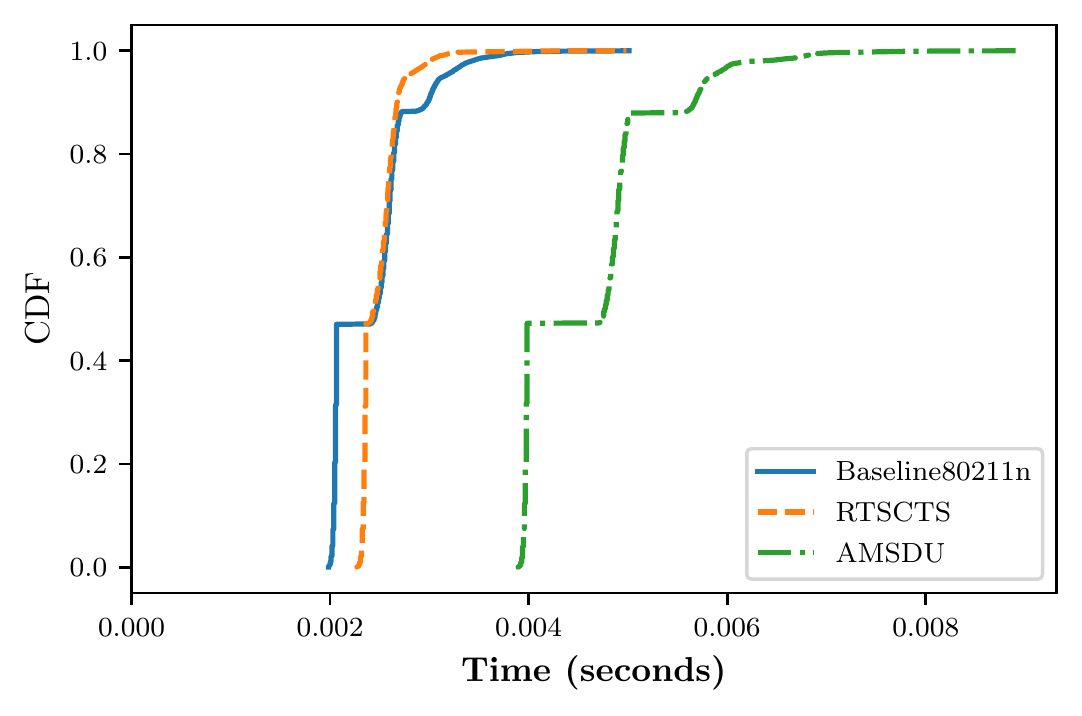}
    \caption{Time-to-empty distributions for the cases of 5 stations and Tx and Rx rates of 144 Mbit/s with different WiFi extensions.}
    \label{fig:time-to-empty-comparison}
\end{figure}

\begin{figure}
    \centering
    \includegraphics[width=\linewidth]{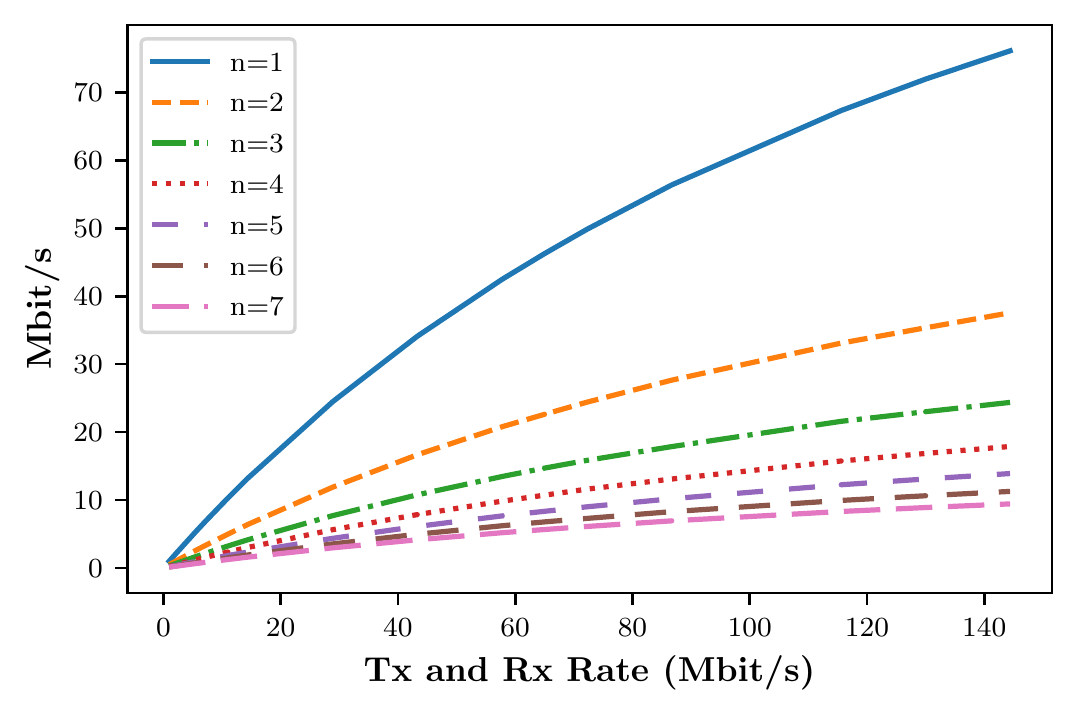}
    \caption{Upper bound on per station throughput for $n$ competing stations with A-MSDU packet aggregation}
    \label{fig:rosoperstaAMSDU}
\end{figure}

We now compare the latency of a WiFi link using packet aggregation to one that does not. Figure \ref{fig:time-to-empty-comparison} shows the time-to-empty CDF for a WiFi network with five stations and transmit and receive rates of 144 Mbit/s. ``Baseline80211n'' is using the exact parameters presented in Table \ref{tab:modernparams}. ``RTSCTS'' uses the RTS/CTS mechanism, and ``AMSDU'' uses packet aggregation with a packet size of 7935 bytes.

\subsubsection{A-MPDU}
Mac Protocol Data Unit aggregation groups several MAC-layer packets into a single transmit opportunity. Once a station has won a transmit opportunity, the station can transmit several packets back-to-back without performing the back-off procedure or waiting for individual ACKs between each MAC-layer packet. The main difference between A-MPDU and A-MSDU is that with A-MPDU, each packet that is part of the aggregate can be ACKed separately. Separate ACKs mean the overhead of packet errors is smaller. Because we are investigating the latency induced by the WiFi DCF specifically, we do not explore the details of A-MPDU, other than to note that the DCF performance will be very similar to that of A-MSDU because both methods compete for transmit opportunities in the same manner. Both methods can send similar amounts of data in a single aggregate.

\subsection{The WiFi performance anomaly}
\label{sec:anomaly}
Heusse et al.\cite{Heusse2003Performance802.11b} first described ``The WiFi Performance Anomaly,'' a phenomenon by which a single low-rate station can lay claim to a large portion of the available airtime. This effect emerges because the WiFi CDF is designed to give each station an equal amount of transmit opportunities. When one station holds on to their transmit opportunity for a longer period than all other stations each time it wins one, the result is an uneven allocation of airtime resources. Heusse et al. show that when this happens, all stations achieve the same throughput as the station using the lowest transmit rate.

\begin{figure}
    \centering
    \includegraphics[width=\linewidth]{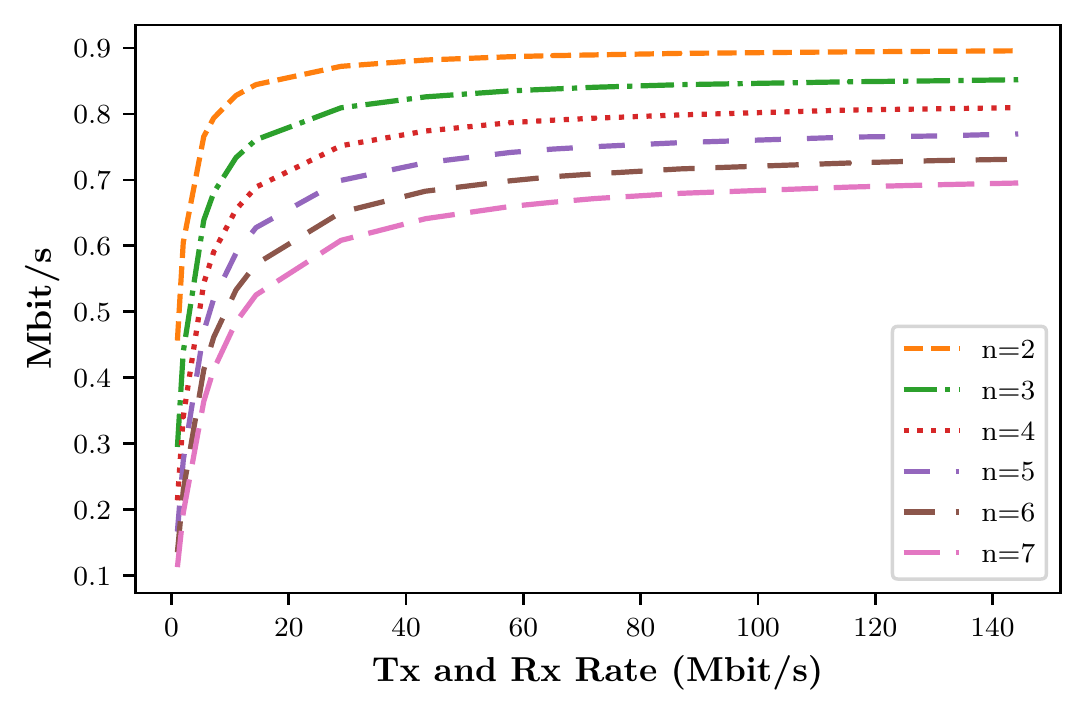}
    \caption{Upper bound on throughput for each of $n$ stations when one of the stations transmit and receive at 1 Mbit/s.}
    \label{fig:anomaly}
\end{figure}

Our WiFi model can readily reproduce this phenomenon. The state representation is modified to include station identifiers such that we can assign different parameters to each station. Then, we assign one station a rate of 1 Mbit/s and vary the rates of the other stations. The resulting throughput per station is shown in figure \ref{fig:anomaly}. As expected, the per-station rate is bounded above by 1 Mbit/s. Our results match those of Heusse et al. \cite{Heusse2003Performance802.11b}.

\section{Conclusion}
\label{sec:conclusion}
In this paper we have presented and validated a novel method for WiFi performance analysis. Our primary contribution is the modeling of complete latency distributions. At the cost of added computational complexity, explicit latency modeling allows more accurate performance analysis by directly modeling the reliability of network outcomes. Our model does this while retaining the ability to produce throughput numbers. We derive upper bounds for WiFi throughput under the requirement that latency and packet loss are bounded. We also investigate the consequences of RTS/CTS and packet aggregation, and reproduce the result known as ``The WiFi performance anomaly''. Our model lets us quantify the impact of trade-offs such as RTS/CTS, where some stability is gained at the expense of added overhead. Using a single framework, we can measure these impacts in terms of throughput and in terms of average latency, jitter and packet loss. This flexible modeling means we can determine which trade-offs should be made depending on the particular use-case of a given WiFi network. Future work will investigate how this new insight can be used to create WiFi control systems that automatically configure the WiFi network to best suit the needs of specific applications, such as video conferencing.

\begingroup
\setlength{\emergencystretch}{.5em}
\printbibliography
\endgroup
\typeout{get arXiv to do 4 passes: Label(s) may have changed. Rerun}
\end{document}